\documentclass[aps,twocolumn,showpacs,superscriptaddress]{revtex4-1}
\usepackage[T1]{fontenc}
\usepackage{amsfonts,amsmath,amssymb,amsbsy}
\usepackage{graphicx,color}
\usepackage{times}
\usepackage[colorlinks=true, linkcolor=blue, citecolor=red, urlcolor=magenta]{hyperref}
\let\mathbf=\boldsymbol
\def\section#1{\medskip\noindent\textbf{#1}\par}
\makeatletter
\renewcommand{\fnum@figure}{\figurename~\textbf{\thefigure}}
\makeatother
\renewcommand{\figurename}{\textbf{Figure}}
\def\blue#1{\textcolor{blue}{#1}}
\def\blue#1{\textcolor{black}{#1}}
\begin{document}

\title{Antiferromagnetic Skyrmion: Stability, Creation and Manipulation}

\author{Xichao Zhang}
\affiliation{Department of Physics, University of Hong Kong, Hong Kong, China}
\affiliation{School of Electronics Science and Engineering, Nanjing University, Nanjing 210093, China}
\author{Yan Zhou}
\email[]{yanzhou@hku.hk}
\affiliation{Department of Physics, University of Hong Kong, Hong Kong, China}
\affiliation{School of Electronics Science and Engineering, Nanjing University, Nanjing 210093, China}
\author{Motohiko Ezawa}
\email[]{ezawa@ap.t.u-tokyo.ac.jp}
\affiliation{Department of Applied Physics, University of Tokyo, Hongo 7-3-1, 113-8656, Japan}

\begin{abstract}\bf\noindent
Magnetic skyrmions are particle-like topological excitations in ferromagnets, which have the topological number $Q=\pm 1$, and hence show the skyrmion Hall effect (SkHE) due to the Magnus force effect originating from the topology.
Here, we propose the counterpart of the magnetic skyrmion in the antiferromagnetic (AFM) system, that is, the AFM skyrmion, which is topologically protected but without showing the SkHE.
Two approaches for creating the AFM skyrmion have been described based on micromagnetic lattice simulations:
(i) by injecting a vertical spin-polarized current to a nanodisk with the AFM ground state;
(ii) by converting an AFM domain-wall pair in a nanowire junction.
It is demonstrated that the AFM skyrmion, driven by the spin-polarized current, can move straightly over long distance, benefiting from the absence of the SkHE.
Our results will open a new strategy on designing the novel spintronic devices based on AFM materials.
\end{abstract}

\date{\today}

\maketitle


\noindent
Skyrmion is a topologically protected soliton in continuous field theory, which is recently realized in both bulk non-centrosymmetric magnetic materials~\cite{Mol,Yu} and thin films~\cite{Heinze}, where the ferromagnetic (FM) background is described by the non-linear sigma model with the Dzyaloshinskii-Moriya interaction (DMI)~\cite{Bogdanov}.
The study of the magnetic skyrmion is one of the hottest topics in condensed matter physics, due to its potential applications in information processing and computing~\cite{SkRev,Fert}.
There are several ways to create magnetic skyrmions, e.g., by applying spin-polarized current to a nanodisk~\cite{Tchoe,Yan}, by applying the laser~\cite{Marco}, from a notch~\cite{Iwasaki} and by the conversion from a domain wall (DW) pair~\cite{NC,XichaoSR2015}.
A magnetic skyrmion can be driven by the spin-polarized current~\cite{Sampaio,XichaoNC2016}.
However, it does not move parallel to the injected current due to the skyrmion Hall effect (SkHE), since its topological number is $\pm 1$.
This will pose a severe challenge for realistic applications which require a straight motion of magnetic skyrmions along the direction of the applied current~\cite{XichaoNC2016}.

\begin{figure}[t]
\centerline{\includegraphics[width=0.425\textwidth]{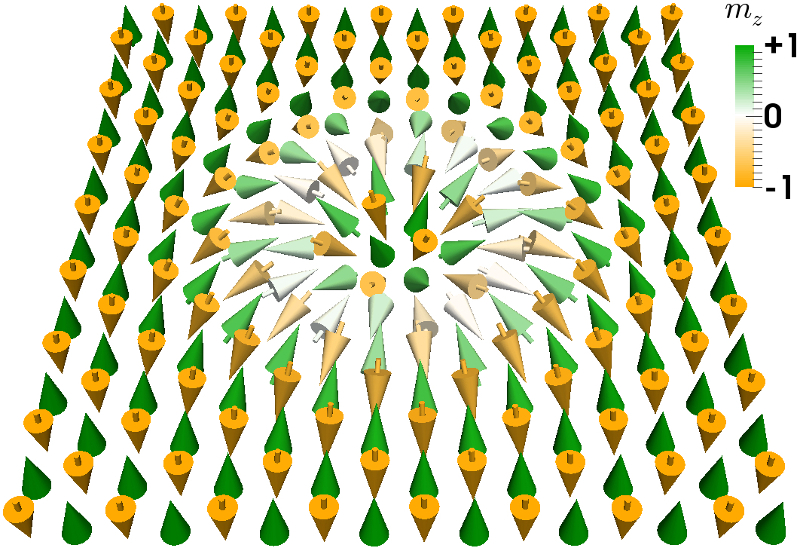}}
\caption{
\textbf{Illustration of an AFM skyrmion spin texture in a 2D AFM monolayer with the two square antiparallel magnetic sublattices.}
The color scale represents the magnetization direction, which has been used throughout this paper: orange is into the plane, green is out of the plane, white is in-plane.
}
\label{FIG1}
\end{figure}

In this work, we demonstrate that a skyrmion can be nucleated in antiferromagnets, as illustrated in Fig.~\ref{FIG1}, based on micromagnetic lattice simulations.
We refer to it as an antiferromagnetic (AFM) skyrmion.
We further show that the AFM skyrmion can move parallel to the applied current since the SkHE is completely suppressed, which is very promising for spintronic applications.

Recently, antiferromagnets emerge as a new field of spintronics~\cite{Nunez,Haney,Helen1,Helen2}.
A one-dimensional topological soliton in antiferromagnets is an AFM DW~\cite{Bode}.
An AFM DW can be moved by spin transfer torque (STT) induced by spin-polarized currents or spin waves~\cite{Ran14,Tveten}.
Besides, a two-dimensional (2D) topological soliton, that is, the magnetic vortex, has been studied in 2D AFM materials~\cite{Baryakhtar}.
The AFM system has an intrinsic two-sublattice structure.
The spins of the ground state are perfectly polarized in each sublattice.
We may calculate the topological number of the spin texture projected to each sublattice.
Hence, we propose to assign a set of topological numbers $(+1, -1)$ to one AFM skyrmion, which shows no SkHE since it has no net magnetization.

We also present two approaches to create an AFM skyrmion.
One is applying a spin-polarized current perpendicularly to a disk region, which flips the spin in the applied region.
The other is a conversion from an AFM DW pair in junction geometry as in the case of the conversion of a FM skyrmion from a FM DW pair~\cite{NC}.
Furthermore, we show that it is possible to move an AFM skyrmion by applying a spin-polarized current.
The AFM skyrmion can travel very long distance without touching the sample edges.
It is also insensible to the external magnetic field.
These results will be important from the applied perspective of magnetic skyrmions.

\vbox{}
\section{Results}

\noindent
\textbf{AFM system.}
We investigate the AFM system with the lattice Hamiltonian,
\begin{equation}
H_{\text{AFM}}=J\sum_{\langle i,j\rangle}\boldsymbol{m}_{i}\cdot\boldsymbol{m}_{j}+\sum_{\langle i,j\rangle}\boldsymbol{D}\cdot(\boldsymbol{m}_{i}\times\boldsymbol{m}_{j})-K\sum_{i}(m_{i}^{z})^{2},
\label{HamilAF}
\end{equation}
where $\boldsymbol{m}_{i}$ represents the local magnetic moment orientation normalized as $|\boldsymbol{m}_{i}|=1$, and $\left\langle i,j\right\rangle$ runs over all the nearest neighbor sites.
The first term represents the AFM exchange interaction with the AFM exchange stiffness $J>0$.
The second term represents the DMI with the DMI vector $\boldsymbol{D}$.
The third term represents the perpendicular magnetic anisotropy (PMA) with the anisotropic constant $K$.

The dynamics of the magnetization $\boldsymbol{m}_{i}$ is controlled by applying a spin current in the current-perpendicular-to-plane (CPP) configuration~\cite{Khv,Sampaio}.
We numerically solve the Landau-Lifshitz-Gilbert-Slonczewski (LLGS) equation,
\begin{align}
\frac{d\boldsymbol{m}_{i}}{dt} =&-|\gamma |\boldsymbol{m}_{i}\times
\boldsymbol{H}_{i}^{\text{eff}}+\alpha \boldsymbol{m}_{i}
\times \frac{d\boldsymbol{m}_{i}}{dt} \notag \\
& +\left\vert \gamma \right\vert \beta (\boldsymbol{m}_{i}\times \boldsymbol{p}
\times \boldsymbol{m}_{i}),
\label{LLGS}
\end{align}
where $\boldsymbol{H}_{i}^{\text{eff}}=-\partial H_{\text{AFM}}/\partial \boldsymbol{m}$ is the effective magnetic field induced by the Hamiltonian equation~(\ref{HamilAF}),
$\gamma$ is the gyromagnetic ratio,
$\alpha$ is the Gilbert damping coefficient originating from spin relaxation,
$\beta$ is the Slonczewski-like STT coefficient,
and $\boldsymbol{p}$ represents the electron polarization direction.
Here, $\beta=|\frac{\hbar }{\mu_{0}e}|\frac{jP}{2dM_{\text{S}}}$ with
$\mu_{0}$ the vacuum magnetic permittivity,
$d$ the film thickness,
$M_{\text{S}}$ the saturation magnetization,
and $j$ the current density.
We take the $\boldsymbol{p}=-\hat{z}$ for creating the AFM skyrmion,
while $\boldsymbol{p}=-\hat{y}$ for moving the AFM skyrmion.
Although an antiferromagnet comprises complex two sublattices of reversely-aligned spins, the STT can be applicable also for the AFM system provided the lattice discreteness effect is taken into account with an ultra-small mesh size in the micromagnetic simulations~\cite{Helen1,Helen2}.
The STT is induced either through spin-polarized current injection from a magnetic tunnel junction polarizer or by the spin Hall effect~\cite{Sampaio,Fino}.
We can safely apply this equation for the AFM system since there is no spatial derivative terms.

A comment is in order.
We cannot straightforwardly use the current-in-plane (CIP) configuration to control the dynamics of the magnetization as it stands,
since spatial derivative terms are involved in the LLGS equation, that is,
$u\boldsymbol{m}_{i}\times\left(\boldsymbol{m}_{i}\times\partial_{x}\boldsymbol{m}_{i}\right)+u^{\prime}\boldsymbol{m}_{i}\times\partial_{x}\boldsymbol{m}_{i}$.

\begin{figure}[t]
\centerline{\includegraphics[width=0.50\textwidth]{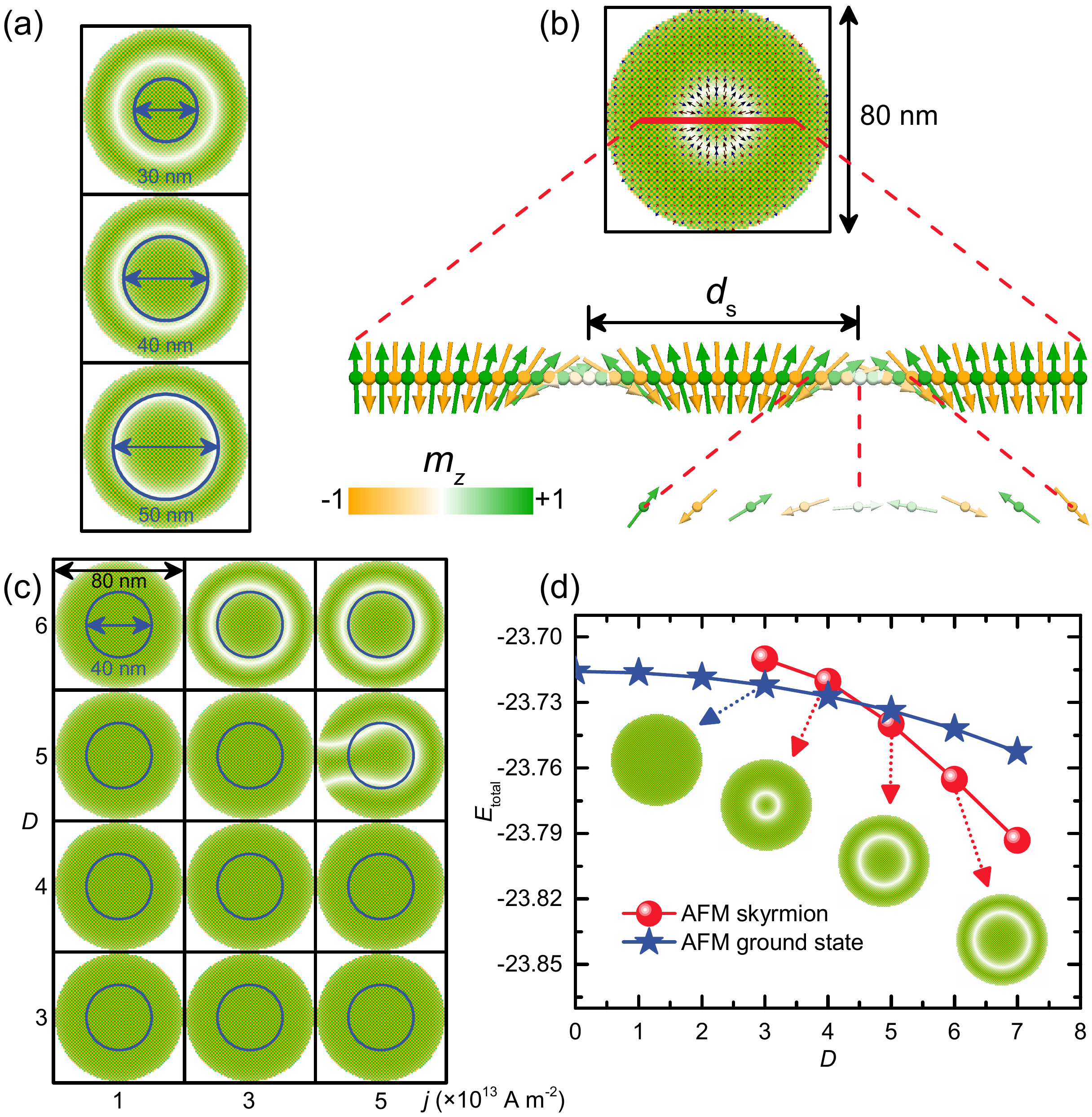}}
\caption{
\textbf{Creation of an isolated AFM skyrmion in a 2D AFM nanodisk using a vertically injected spin-polarized current.}
(\textbf{a}) AFM skyrmion in the nanodisk with a diameter of $80$ nm created by a $2$-ns-long spin-polarized current pulse ($j=5\times 10^{13}$ A m$^{-2}$) perpendicularly injected into the nanodisk in a circle region with different size followed by a $1$-ns-long relaxation (\blue{see Supplementary Movie 1}).
The spin-polarized current injection region is denoted by the blue circle, of which the diameter equals $30$ nm, $40$ nm, $50$ nm, respectively.
The size of all AFM skyrmions is found to be identical irrelevant of the current injection region.
(\textbf{b}) Magnetization distribution of the AFM skyrmion in an AFM nanodisk.
It is made of a toroidal DW with fixed radius and width determined by the material parameters.
The AFM skyrmion size $d_{\text{s}}$ is defined by the diameter of the white circle, where $m_{z}=0$.
(\textbf{c}) A $2$-ns-long spin-polarized current pulse with different current density $j$ is perpendicularly injected into the nanodisk with a diameter of $80$ nm followed by a $1$-ns-long relaxation.
The initial state of the nanodisk is the AFM ground state.
The spin-polarized current injection region is denoted by the blue circle, of which the diameter equals $40$ nm.
(\textbf{d}) Total micromagnetic energy $E_{\text{total}}$ for an isolated AFM skyrmion and the AFM ground state as a function of the DMI constant $D$.
The $D$ and $E_{\text{total}}$ are in unit of $10^{-21}$ J and $10^{-17}$ J, respectively.
}
\label{FIG2}
\end{figure}

\begin{figure}[t]
\centerline{\includegraphics[width=0.50\textwidth]{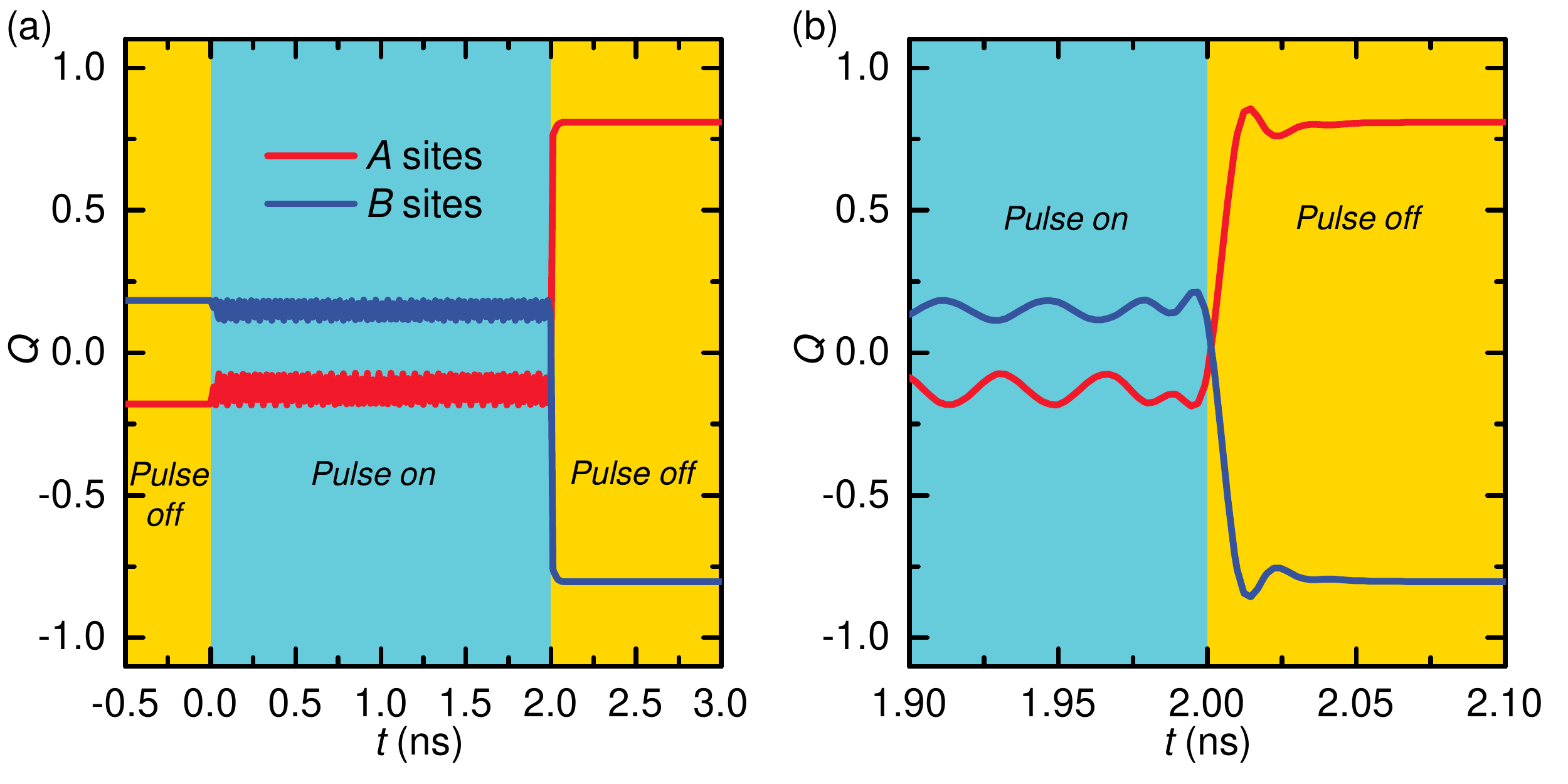}}
\caption{
\textbf{The time evolution of the topological number $Q$, that is, the skyrmion number, in the nucleation process of an isolated AFM skyrmion in a 2D AFM nanodisk with the two square sublattices of \textit{A} sites and \textit{B} sites.}
(\textbf{a}) A $2$-ns-long spin-polarized current pulse ($j=3\times 10^{13}$ A m$^{-2}$) perpendicularly injected into an AFM nanodisk with a diameter of $80$ nm in a circle region with a diameter of $40$ nm followed by a $1$-ns-long relaxation.
(\textbf{b}) Close-up of the time evolution shown in (\textbf{a}), where the topological number of the two sublattices significantly changes from $\sim 0$ to $\pm 1$.
}
\label{FIG3}
\end{figure}

\vbox{}
\noindent
\textbf{Topological stability.}
The skyrmion carries the topological number.
In the continuum theory it is given by
\begin{equation}
Q=-{\frac{1}{4\pi}}\int d^{2}x\boldsymbol{m}(\boldsymbol{x})\cdot\left(\partial_{x}\boldsymbol{m}(\boldsymbol{x})\times\partial_{y}\boldsymbol{m}(\boldsymbol{x})\right).
\label{PontrNumbe}
\end{equation}
However, the AFM system has a two-sublattice structure made of the \textit{A} and \textit{B} sublattices.
In our numerical computation we employ the discretized version of the topological charge equation~(\ref{PontrNumbe}),
\begin{equation}
Q_{\tau}=-{\frac{1}{4\pi}}\sum_{ijk}\boldsymbol{m}_i^{\tau}\cdot\left(\boldsymbol{m}_j^{\tau}\times\boldsymbol{m}_k^{\tau}\right),
\label{PontrDiscr}
\end{equation}
for each sublattice ($\tau=A, B$).
Hence, we propose to assign a set of two topological numbers $(Q_{\textit{A}}, Q_{\textit{B}})$ to one skyrmion.
We obtain $Q_{\textit{A}}=-Q_{\textit{B}}=1$ for a skyrmion in a sufficiently large area.
Even if the skyrmion spin texture is deformed, its topological number cannot change.
A skyrmion can be neither destroyed nor separated into pieces, that is, it is topologically protected.

\vbox{}
\noindent
\textbf{Creation of an AFM skyrmion by a vertical spin current.}
We employ a CPP injection with a circular geometry in a nanodisk.
The CPP injection induces spin flipping at the current-injected region.
When we continue to apply the current, the spins continue to flip.
As soon as we stop the current, an AFM skyrmion is nucleated to lower the DMI and AFM exchange energies (\blue{see Supplementary Movie 1}).
It is relaxed to the optimized radius irrespective of the injected region, as shown in Fig.~\ref{FIG2}a (\blue{see Supplementary Movie 2}).
Once it is relaxed, it stays as it is for long, demonstrating its static stability.
We show the spin configuration of an AFM skyrmion obtained numerically in Fig.~\ref{FIG2}b.
It is made of a toroidal DW with fixed radius and width determined by the material parameters.
There exists a threshold current density to create an AFM skyrmion, as shown in Fig.~\ref{FIG2}c.
It is natural that the spins cannot be flipped if the injected current is not strong enough.

The time-evolution of the topological charges of the AFM system is shown in Fig.~\ref{FIG3}.
Note that there is a non-zero topological number $Q_{\tau}^0$ in the AFM background state, which is created by the tilting magnetization at edges due to the DMI.
It is $Q_A^0=-0.1827$ for the \textit{A} sites.
The topological charge oscillates during the CPP injection.
As soon as the CPP injection is off, the topological charge develops suddenly to a fixed values.
By subtracting $Q_A^0$ from that of the AFM skyrmion in the \textit{A} sites, we find $Q_A=0.9865$, which is almost $1$.
Similarly, we find $Q_B$ is almost $-1$.

The AFM skyrmion can be created equally by a vertical current injection polarized along the $+z$-direction or the $-z$-direction (\blue{see Supplementary Movie 3}).

\begin{figure}[t]
\centerline{\includegraphics[width=0.50\textwidth]{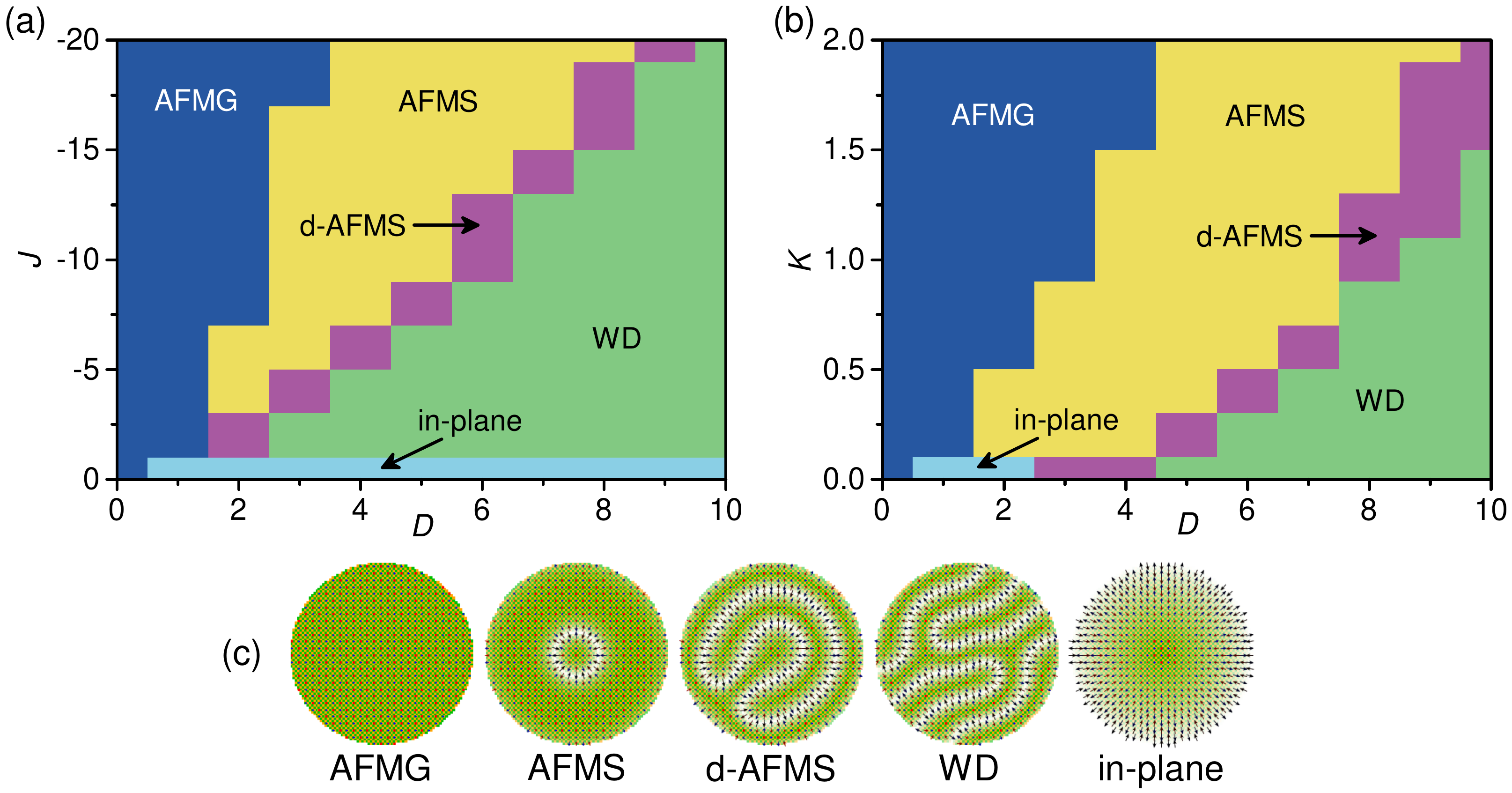}}
\caption{
\textbf{Phase diagrams of the AFM skyrmion as functions of the exchange stiffness $J$ and the DMI constant $D$ (\textbf{a}), and as functions of the PMA constant $K$ and $D$ (\textbf{b}).}
An AFM skyrmion cannot exist in the AFM ground (AFMG) state region (blue color).
An AFM skyrmion can exist as a static stable object in the AFM skyrmion (AFMS) state region (yellow color).
An AFM skyrmion is distorted to lower the DMI energy in the distorted AFM skyrmion (d-AFMS) state region (violet color).
The ground state contains worm domains in the worm domain (WD) state region (green color).
All spins are almost in plane because the exchange interaction or the anisotropy strength is too weak in the in-plane state region (light blue color).
(\textbf{c}) Illustration of the states shown in the phase diagrams.
The simulation is based on the nanodisk with a diameter of $80$ nm.
The $J$, $D$ and $K$ are in unit of $10^{-21}$ J.
The default values of $J$ and $K$ equal $15\times 10^{-21}$ J and $0.8\times 10^{-21}$ J, respectively.
}
\label{FIG4}
\end{figure}

\vbox{}
\noindent
\textbf{Phase diagrams.}
A skyrmion is topologically protected.
Nevertheless, it may shrink or expand with the topological charge unchanged.
We present a phase diagram in Fig.~\ref{FIG4}.
It is convenient to understand it in terms of the DMI constant $D$.
The DMI prevents a skyrmion from shrinking in antiferromagnets as in the case of ferromagnets.
(1) Near $D=0$, a skyrmion shrinks and disappears (blue region).
(2) There are two cases when a skyrmion exists as a static stable object (yellow region): see also Fig.~\ref{FIG2}d.
In one case (smaller $D$), its energy is more than that of the AFM ground state.
It is an energetically metastable state, but it is topologically stable.
In the other case (larger $D$), its energy is less than that of the AFM ground state.
It would undergo condensation if it were not topologically protected.
(3) When $D$ becomes larger, a skyrmion is distorted to reduce the DMI energy, which we call a distorted AFM skyrmion (violet region).
(4) When $D$ becomes sufficiently large, a deformed skyrmion touches the edge, forming worm domains (green region).

\begin{figure}[t]
\centerline{\includegraphics[width=0.50\textwidth]{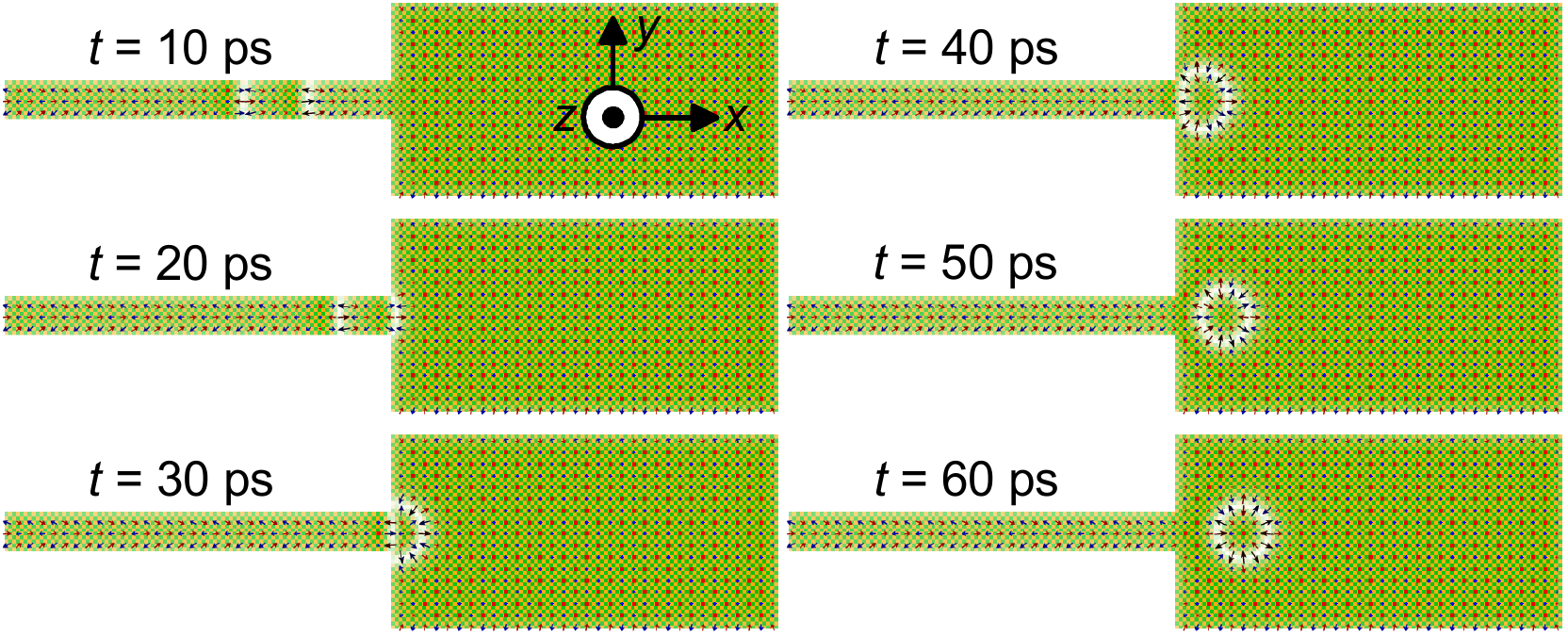}}
\caption{
\textbf{Creation of an isolated AFM skyrmion in a 2D AFM nanotrack via AFM DW pair driven by a vertical spin-polarized current.}
A vertical current with a density of $j=4.5\times 10^{11}$ A m$^{-2}$ in the wide part is applied perpendicular to the nanotrack from the bottom.
An AFM skyrmion is created from an AFM DW pair driven by the current moving from the narrow part to the wide part of the nanotrack, $D=4\times 10^{-21}$ J (\blue{see Supplementary Movie 4}).
The current density inside the wide part of the nanotrack is proportional to the current density inside the narrow part of the nanotrack with respect to the ratio of the narrow width ($20$ nm) to the wide width ($100$ nm).
}
\label{FIG5}
\end{figure}

\vbox{}
\noindent
\textbf{Creation of an AFM skyrmion from an AFM DW wall pair.}
A FM skyrmion can be created from a FM DW pair using a junction geometry~\cite{NC}.
We show that a similar mechanism works in creating the AFM skyrmion as shown in Fig.~\ref{FIG5}.
We first make an AFM DW pair through the CPP injection with $\boldsymbol{p}=-\hat{z}$.
The AFM DW pair is shifted by applying a spin-polarized current through the STT on AFM DW~\cite{RanL} as shown in Fig.~\ref{FIG5} (see the process from $t=10$ ps to $t=20$ ps in Fig.~\ref{FIG5}).
Here we consider the vertical injection of a spin current polarized along the $-y$-direction.
The CPP injection moves the AFM DW in the rightward direction ($+\hat{x}$).
When the AFM DW arrives at the junction interface ($t=20$ ps), both the end spins of the DW are pinned at the junction, whereas the central part of the DW continues to move due to STT in the wide part of the nanotrack.
Therefore, the structure is deformed into a curved shape and an AFM skyrmion texture forms at $t=30$ ps.
This process is analogous to blowing air through soapy water using bubble pipes or plastic wands to create soap bubbles.
The skyrmion will break away from the interface when the bulk of its structure continues to move rightward as shown at $t=40$ ps.
By continuously ``blowing'' AFM DWs into the junction, a train of AFM skyrmions is generated.

\begin{figure}[t]
\centerline{\includegraphics[width=0.485\textwidth]{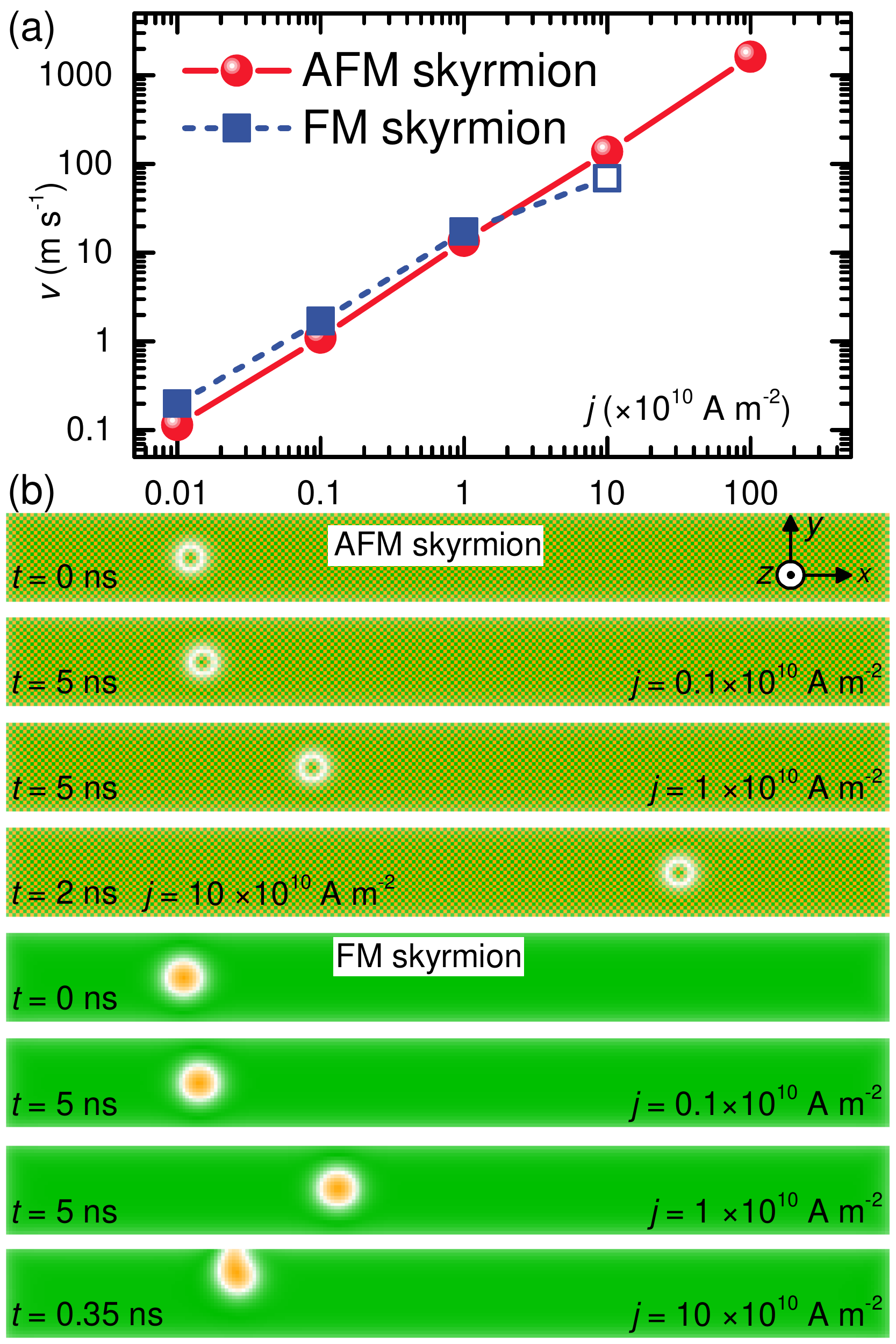}}
\caption{
\textbf{Motion of an isolated AFM (FM) skyrmion in a 2D AFM (FM) nanotrack driven by a vertically injected spin-polarized current.}
(\textbf{a}) AFM skyrmion velocity and FM skyrmion velocity as functions of current density $j$ with the CPP geometry.
For the AFM skyrmion case, the current is applied along $+\hat{z}$ but polarized along $-\hat{y}$, $J=15\times 10^{-21}$ J.
For the FM skyrmion case, the current is applied along $+\hat{z}$ but polarized along $+\hat{y}$, $J=-15\times 10^{-21}$ J.
For both cases, $\protect\alpha=0.3$, $D=3.5\times 10^{-21}$ J, $K=0.8\times 10^{-21}$ J.
The open symbol denotes the destruction of the FM skyrmion due to the SkHE.
(\textbf{b}) Top-views of vertical current-driven AFM skyrmion and FM skyrmion at selected current densities and times.
}
\label{FIG6}
\end{figure}

\vbox{}
\noindent
\textbf{Current-driven motion of an AFM skyrmion in a nanotrack.}
We can move the AFM skyrmion by the CPP injection as in the case of the FM skyrmion.
We show the relation between the magnitude of the injected current and the velocity in Fig.~\ref{FIG6}a, where the velocity is proportional to the injected current.

We recall that the FM skyrmion is easily destroyed by touching the sample edges due to the SkHE.
At the same time, the maximum velocity of the FM skyrmion in a FM nanotrack is typically much less than $10^{2}$ m s$^{-1}$, limited by the confining force of $\sim (D^{2}/J)$~\cite{IwasakiNL}.

Conversely, there is no SkHE for the AFM skyrmion.
Hence, it can move straightly in an AFM nanotrack without touching the edge.
It is shown in \blue{Supplementary Movie 5}, where a chain of encoded AFM skyrmions moves in a nanotrack with a speed of $\sim 1700$ m s$^{-1}$ driven by a vertical current without touching edges.

In Fig.~\ref{FIG6}b we compare the AFM and FM skyrmions.
The velocity of AFM skyrmions can be very large compared to FM skyrmion, which is suitable for ultrafast information processing and communications.
The steady motion of AFM skyrmions is demonstrated in \blue{Supplementary Movie 6}, where they move in a thin film without boundary effect driven by the vertical spin current.
This highly contrasts with the case of FM skyrmions demonstrated in \blue{Supplementary Movie 7}, where skyrmions do not move either parallel or perpendicular to the film edges.

\section{Discussion}

\noindent
We have proposed magnetic skyrmions in the AFM system.
The dynamics of AFM skyrmions is very different from those in the FM system, since they are topologically protected and are free from the SkHE.
We have first checked that our simulation software reproduces a linear dispersion relation inherent to the two-sublattice structure of the AFM system, and then employ it to explore various properties of the AFM skyrmion.
It is worth mentioning there are two recent preprints~\cite{CYM,TretiakovAFMSkyrmion} on AFM skyrmions, including the preliminary version of the present work~\cite{CYM}. Our work is focused on the injection and vertical spin current-driven dynamics of AFM skyrmions.
Regarding the other work in Ref.~\onlinecite{TretiakovAFMSkyrmion}, the thermal properties as well as in-plane current-induced dynamics of an AFM skyrmion have been studied and a high-speed motion ($v\sim 10^{3}$ m s$^{-1}$) of an AFM skyrmion has also been shown in the absence of the SkHE, consistent with the present work.
We believe that the AFM skyrmions will play a very significant role in the emerging field of AFM spintronics.

\section{Methods}

\noindent
\textbf{Modeling and simulation.}
We perform the micromagnetic simulations using the Object Oriented MicroMagnetic Framework (OOMMF) together with the DMI extension module~\cite{Sampaio,Bou,Dona,Rohart,XichaoSR2014}.
The time-dependent magnetization dynamics is governed by the LLGS equation~\cite{Brown,Gilbert,LL,Thia,Thia2}.
The OOMMF has been developed originally and used extensively for the simulation of FM systems, and we have checked that one may use it to analyze the nanotexture in the AFM system as well.
Indeed, we have successfully reproduced a linear dispersion relation inherent to the two-sublattice structure, as shown in \blue{Supplementary Figure 1}.

For micromagnetic simulations, we consider $0.4$-nm-thick magnetic nanodisks and nanotracks on the substrate.
With respect to the material parameters, we recall~\cite{Kittel} that an antiferromagnet is a special case of a ferrimagnet for which both sublattices \textit{A} and \textit{B} have equal saturation magnetization.
Both the DMI and the PMA arise from the spin orbit coupling, albeit in different forms.
We have checked that our results hold for a wide range of material parameters (\blue{cf. Fig.~\ref{FIG4}}).
Here, we use the parameters of the same order as those given in Ref.~\onlinecite{Saiki} for AFM materials.
We thus adopt the magnetic parameters from Refs.~\onlinecite{Fert,Sampaio}:
the Gilbert damping coefficient $\alpha=0.3$,
the gyromagnetic ratio $\gamma=-2.211\times 10^{5}$ m A$^{-1}$ s$^{-1}$,
the sublattice saturation magnetization $M_{\text{S}}=290$ kA m$^{-1}$,
the exchange constant $J=0\sim 20\times 10^{-21}$ J,
the DMI constant $D=0\sim 10\times 10^{-21}$ J,
and the PMA constant $K=0\sim 2\times 10^{-21}$ J unless otherwise specified.
All samples are discretized into tetragonal cells of $1$ nm $\times$ $1$ nm $\times$ $0.4$ nm in the simulation, which ensures reasonable numerical accuracy as well as run time.
The output time step is fixed at $1$ ps for the simulation of the dispersion relation, which is increased to $10$ ps for the simulation of the skyrmion dynamics.
The polarization rate of the spin-polarized current is defined as $P=0.4$ in all simulations.
The Zeeman field is set as zero because the AFM skyrmion, having no net magnetization, is insensitive to it (\blue{see Supplementary Figure 2}).



\section{Acknowledgements}
\noindent
M.E. is very much grateful to N. Nagaosa for many helpful discussions on the subject.
X.Z. thanks J. Xia for useful discussions.
X.Z. was supported by JSPS RONPAKU (Dissertation Ph.D.) Program and was partially supported by the Scientific Research Fund of Sichuan Provincial Education Department (Grant No.~16ZA0372).
Y.Z. acknowledges the support by National Natural Science Foundation of China (Project No.~1157040329), the Seed Funding Program for Basic Research and Seed Funding Program for Applied Research from the HKU, ITF Tier 3 funding (ITS/203/14), the RGC-GRF under Grant HKU 17210014, and University Grants Committee of Hong Kong (Contract No.~AoE/P-04/08).
M.E. thanks the support by the Grants-in-Aid for Scientific Research from MEXT KAKENHI (Grants No.~25400317 and No.~15H05854).

\section{Author Contributions}
\noindent
M.E. conceived the project. Y.Z. coordinated the project. X.Z. carried out the numerical simulations supervised by Y.Z. All authors interpreted the results and prepared the manuscript and supplementary information.

\section{Additional Information}
\noindent
Supplementary information is available in the online version of the paper.
Correspondence and requests for materials should be addressed to M.E. and Y.Z.

\section{Competing Financial Interests}
\noindent
The authors declare no competing financial interests.

\end{document}